\documentclass[twocolumn,aps,preprintnumbers,amssymb,amsmath,superscriptaddress,nofootinbib]{revtex4}
\pdfoutput=1
\usepackage{amsmath,amssymb,euscript}
\usepackage{dcolumn} 
\usepackage{bm}
\usepackage{slashed}
\usepackage{color}
\usepackage{accents}
\usepackage{hyperref}
\usepackage{ulem}
\usepackage{epsfig}
\usepackage{varioref}
\usepackage{xcolor}
\usepackage{verbatim}
\usepackage{graphicx}

\def\beq{\begin{equation}}
\def\eeq{\end{equation}}

\renewcommand{\emph}{\textit}

\graphicspath{{figs/}}

\begin{document}
\onecolumngrid\begin{flushright} P3H-19-043 \end{flushright}\twocolumngrid

\title{Probing dark sectors with long-lived particles at Belle II}

\author{Anastasiia Filimonova, Ruth Sch\"afer, Susanne Westhoff\\
\textit{ \small{Institute for Theoretical Physics, Heidelberg University, D-69120 Heidelberg, Germany} }}

\vspace{1.0cm}
\begin{abstract}
\vspace{0.2cm}\noindent We propose a new search for light scalar singlets in rare meson decays. For couplings well below the electroweak interaction strength, the scalar is long-lived at detector scales and decays into displaced pairs of leptons or light mesons. We show that Belle II has a remarkable potential to probe scalars in the GeV range with  couplings as small as $10^{-5}$. The predicted sensitivity is higher than at the long-baseline experiments NA62 and FASER and comparable with projections for FASER 2. We also investigate signatures of invisibly decaying scalars in rare meson decays with missing energy.
\end{abstract}
\maketitle


\section{Introduction}
\noindent
It could well be that the Higgs boson is not the only scalar in nature. A second scalar that mixes with the Higgs boson can be realized in minimal renormalizable extensions of the standard model~\cite{Patt:2006fw,OConnell:2006rsp}. Such scalars could naturally be light~\cite{Piazza:2010ye}, which offers attractive solutions to big open questions in particle physics and cosmology. For instance, light scalars are thermal dark matter candidates~\cite{Silveira:1985rk} or mediators to a dark sector~\cite{Pospelov:2007mp}, could generate the electroweak hierarchy through cosmological relaxation~\cite{Graham:2015cka}, facilitate  baryogenesis~\cite{Espinosa:1993bs,Profumo:2007wc,Croon:2019ugf}, or play the role of an instanton during inflation~\cite{Bezrukov:2009yw}.

Extensive searches for new scalars at particle colliders and fixed-target experiments have probed  couplings of $1 - 10^{-3}$ over a wide mass range up to the electroweak scale. Ref.~\cite{Winkler:2018qyg} gives a comprehensive overview. Complementary to collider searches, astrophysical and cosmological observations are sensitive to very weak couplings below $10^{-7}$ for scalars around the GeV scale and set strong bounds on sub-GeV scalars~\cite{Fradette:2018hhl,Bondarenko:2019vrb}.

In this work, we focus on scalars in the GeV range which can be resonantly produced in $B$ and $K$ meson decays through loop-induced flavor-changing neutral currents~\cite{Grinstein:1988yu,Chivukula:1988gp,Batell:2009jf}. The phenomenology of light scalars in  meson decays like $B\to K\mu\bar\mu$, $B_s \to \mu\bar\mu$, or $K\to \pi \mu\bar\mu$ has been explored for instance in Refs.~\cite{Schmidt-Hoberg:2013hba,Dolan:2014ska,Krnjaic:2015mbs,Dev:2017dui,Boiarska:2019jym}. We show that Belle II can search for \emph{displaced} meson decays and thus penetrate an unexplored territory of scalar couplings in the range $10^{-3} - 10^{-5}$. The key to this new search is that scalars with such tiny couplings are long-lived at detector scales, leaving traces of displaced vertices from their decay products~\cite{Batell:2009jf,Clarke:2013aya}. We predict that Belle II has a larger reach than the long-baseline experiments NA62 and FASER and competes with searches for long-lived particles at the proposed dedicated experiments FASER\,2, CODEX-b, SHiP or MATHUS\-LA~\cite{Beacham:2019nyx}.

Complementary to displaced decays, we investigate invisible decays of scalars in rare meson decays. Invisible decays are particularly relevant in the context of dark matter, where the scalar is the mediator of a new force between standard-model particles and a dark sector~\cite{Bird:2004ts,Badin:2010uh,Krnjaic:2015mbs,Dev:2017dui,Evans:2017kti,Chen:2018vkr}. We predict that Belle II can improve the current sensitivity to invisibly decaying scalars with a dedicated search for two-body decays $B\to K\slashed{E}$ with missing energy in the final state.

After reviewing the phenomenology of light scalars in meson decays in Sec.~\ref{sec:meson-decays}, we  discuss signatures with missing energy in Sec.~\ref{sec:inv}. In Sec.~\ref{sec:dv} we make predictions for displaced meson decays at Belle II and finally summarize our main results in Sec.~\ref{sec:conclusions}.

\section{Dark scalars in meson decays}\label{sec:meson-decays}
\noindent
We extend the standard model by a real scalar field $\phi$ and a Dirac fermion $\chi$, both being singlets under the strong and electroweak forces. The fermion is charged under a discrete $\mathbb{Z}_2$ symmetry, so that it does not mix with neutrinos and is a stable dark matter candidate. The new interactions and mass terms are described by the Lagrangian
\begin{align}
	\!\mathcal{L}&= - \tfrac{1}{2} m_{\phi}^2\phi^2 - \lambda_3\,|H|^2\phi - y_\chi \bar \chi\chi \phi - \tfrac{1}{2}m_{\chi}\bar\chi\chi\,.
\end{align}
The scalar mediates a new force between the Higgs field $H$ and the dark fermion, which represents here a potentially more complex dark sector. We neglect a possible quartic interaction $|H|^2\phi^2$. After electroweak symmetry breaking the scalar $\phi$ mixes with the neutral component of the Higgs field into a dark scalar $S$ and the observed 125-GeV Higgs boson $h$. 
The fermion couplings of the physical scalars are now given by
\begin{align}
	\mathcal{L}_y &= y_\chi \left(s_\theta \, \bar\chi\chi h - c_\theta \, \bar\chi\chi S \right)\\\nonumber
	&\phantom{=\ }- \sum_f \frac{m_f}{v} \left ( c_\theta \, \bar ff h + s_\theta \, \bar ff S \right),
\end{align}
where $s_\theta$ and $c_\theta$ denote the sine and cosine of the mixing angle $\theta$, $m_f$ is the fermion mass and $v = 246\,\text{GeV}$ is the vacuum expectation value of the Higgs field. For technical details we refer the reader to Appendix~\ref{sec:app}. The dark scalar inherits the flavor-hierarchical Yukawa couplings of the Higgs boson to standard-model fermions $f$. This flavor hierarchy is characteristic for scalar mediators and distinguishes them from mediators with flavor-universal couplings, for example dark photons coupling through kinetic mixing~\cite{Holdom:1985ag}.

The phenomenology of the dark scalar critically depends on its decay width
\begin{align}\label{eq:scalar-lifetime}
    \Gamma_S = s_\theta^2\,\Gamma_{\rm SM} + c_\theta^2\,\Gamma_{\chi\bar{\chi}}\,,
\end{align}
where $\Gamma_{\rm SM}$ and $\Gamma_{\chi\bar\chi}$ denote the partial widths into standard-model particles and dark fermions, and we have factored out the dependence on the mixing angle $\theta$. The branching ratios into leptons $\ell$ and dark fermions $\chi$ are
\begin{align}
   \!\!\!\mathcal{B}(S\to \ell\bar \ell) & = \frac{s_\theta^2\,\Gamma_{\ell\bar \ell}}{\Gamma_S} = \frac{m_\ell^2 s_\theta^2}{8 \pi v^2} \frac{m_S}{\Gamma_S} \bigg(1-\frac{4m_\ell^2}{m_S^2}\bigg)^{3/2}\!,\label{eq:partial}\\\nonumber
   \!\!\!\mathcal{B}(S\to \chi\bar\chi) & = \frac{c_\theta^2\,\Gamma_{\chi\bar \chi}}{\Gamma_S} = \frac{y_\chi^2 c_\theta^2}{8\pi}\frac{m_S}{\Gamma_S} \bigg(1-\frac{4m_\chi^2}{m_S^2}\bigg)^{3/2}\!.
\end{align}
Scalars in the GeV range decay into ``visible'' final states with leptons or light mesons. For hadronic decays we adopt the predictions from Ref.~\cite{Winkler:2018qyg}, which are based on dispersion relations for $m_S < 2\,\text{GeV}$ and on a perturbative spectator model for higher masses. Below the di-muon threshold the scalar decays into electrons or photons~\cite{Fradette:2018hhl}. 

Visible decays into standard-model particles dominate for $m_S < 2m_\chi$, where the decay to dark fermions is kinematically forbidden. Invisible decays dominate for $m_S > 2m_\chi$ and $y_\chi c_\theta > m_\ell s_\theta/v$. Due to the small Yukawa coupling $m_\ell/v$, this condition is fulfilled even for very weak dark fermion couplings $y_\chi$.

The production of dark scalars in $B_q$ meson decays relies on effective flavor-changing currents ($q=s,d$)
\begin{align}
	\mathcal{L}_{\text{eff}}&= \frac{C_{bq}}{v}\left(m_b\,\overline{q}_L b_R + m_q\,\overline{q}_R b_L\right)S\,.
\end{align}
Since the fundamental scalar couplings are flavor-diagonal and hierarchical, these interactions are loop-induced through the large top-quark coupling. The Wilson coefficient
\begin{align}
	C_{bq} & = \frac{3\sqrt{2}G_F m_t^2}{16\pi^2}V_{tb}V_{tq}^\ast\,s_\theta + \mathcal{O}\left(\frac{m_S^2}{m_W^2}\right)
\end{align}
is identical to the Higgs penguin~\cite{Grzadkowski:1983yp,Willey:1982mc}, multiplied by the scalar mixing $s_\theta$. This interaction induces two-body decays $B\to MS$, where $M=K,K^\ast,\pi,\rho, \dots$, or $K\to \pi S$ provided that the final state can be produced resonantly. The branching ratio for $B^+\to K^+S$ decays is given by~\footnote{For general expressions for $B\to M S$ and $K\to MS$ decays see Refs.~\cite{Kamenik:2011vy,Boiarska:2019jym}.}
\begin{align}
	& \mathcal{B}(B^+\to K^+ S) = \frac{\sqrt{2}G_F\left|C_{bs}\right|^2}{64\pi \Gamma_{B^+} m_B^3} \frac{(m_b+m_s)^2}{(m_b - m_s)^2}\, f_0^2\big(m_S^2\big)\\\nonumber
	& \, \times \big(m_B^2-m_K^2\big)^2\big[(m_B^2-m_K^2-m_S^2)^2-4m_K^2m_S^2\big]^{\frac{1}{2}},
\end{align}
where $\Gamma_{B^+}$ is the total decay width of the $B^+$ meson and $f_0(m_S^2)$ is the scalar hadronic form factor at momentum transfer $q^2 = m_S^2$~\cite{Bailey:2015dka}. With $\mathcal{B}(B^+\to K^+S)\approx 0.5\,s_\theta^2$ the scalar production rate is large for sizeable mixing. Observable branching ratios for $B\to K\ell\bar\ell$ and $B\to K\chi\bar\chi$ decays through a narrow scalar resonance are finally given by
\begin{align}\label{eq:B-to-KS}
	\mathcal{B}\big(B\to K S\big)\mathcal{B}\big(S\to \ell\bar \ell\big) & \propto s_\theta^2\frac{s_\theta^2\Gamma_{\ell\bar \ell}}{\Gamma_S}\,,\\\nonumber
		\mathcal{B}\big(B\to K S\big)\mathcal{B}\big(S\to\chi\bar\chi\big) & \propto  s_\theta^2\frac{c_\theta^2\Gamma_{\chi\bar \chi}}{\Gamma_S}\,.
\end{align}
For $\mathcal{B}(S\to \chi\bar\chi) = 0$ the scalar always decays into visible final states. If the mixing is small the scalar becomes long-lived at detector scales and leaves signatures with displaced vertices, for instance displaced muon pairs from $B\to K S(\to \mu\bar\mu)$ decays. For $\mathcal{B}\big(S\to\chi\bar\chi\big) \approx 1$ the scalar decays invisibly and creates signatures with missing energy.

\section{Missing energy signatures}\label{sec:inv}
\noindent
We start by exploring observables with missing energy, assuming that the scalar decays dominantly into invisible final states. Rare meson decays like $B\to K \slashed{E}$ are very sensitive to scalar contributions $B\to KS(\to \chi\bar\chi)$. From Eq.~\eqref{eq:B-to-KS} we see that for $\Gamma_S \approx c_\theta^2 \Gamma_{\chi\bar\chi}$ the rate scales like $s_\theta^2$ and does not depend on the scalar's decay. Searches for $B\to K\slashed{E}$ are therefore blind to the exact properties of the dark sector and applicable for a wider range of models.

\begin{figure}[t!]
	\centering
	\includegraphics[width=\linewidth]{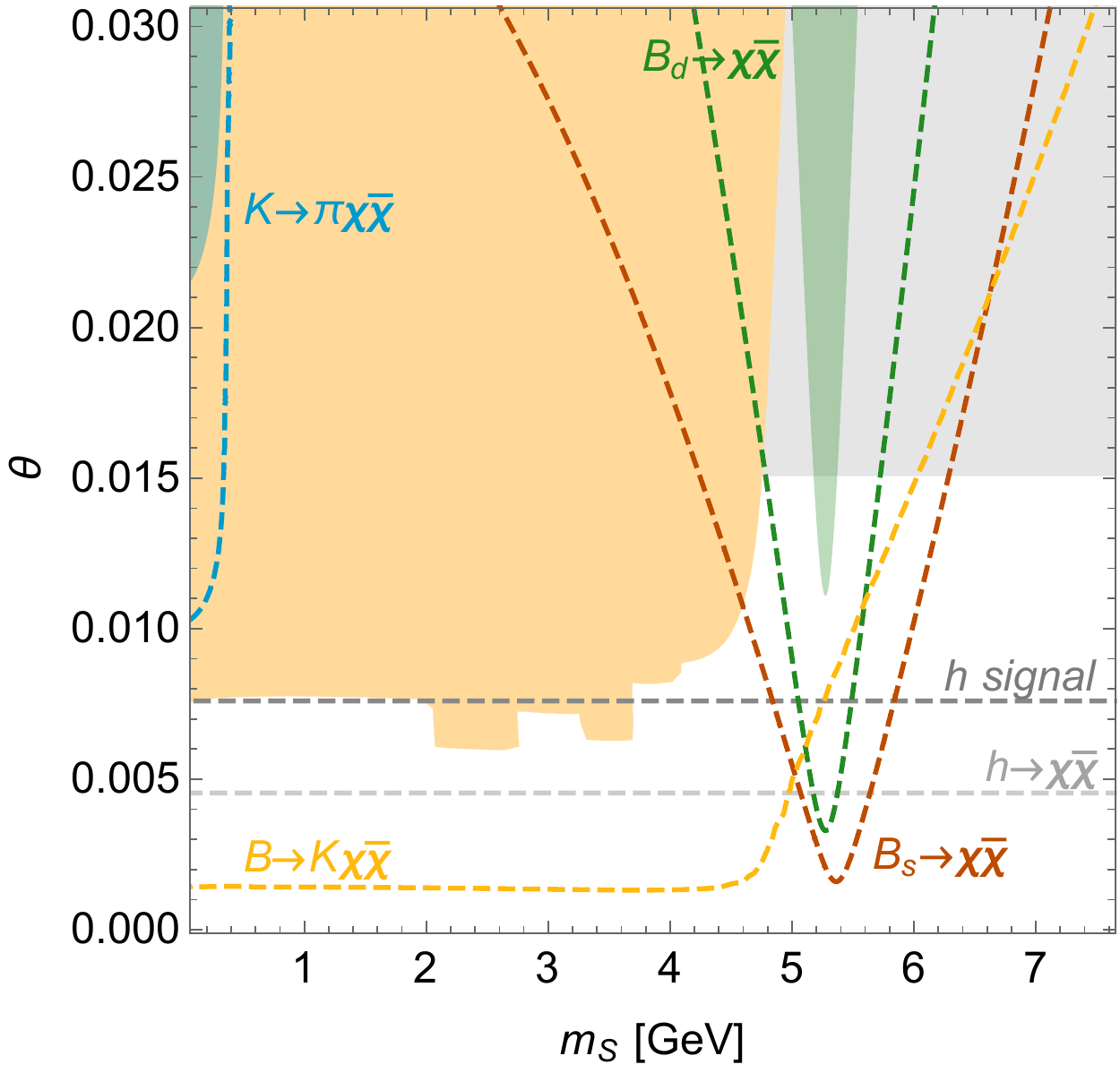}
	\caption{Searches for invisibly decaying dark scalars. Shaded regions are excluded by searches for $B\to K\slashed{E}$~\cite{Lees:2013kla} (yellow) at BaBar and $B_d \to \slashed{E}$~\cite{Hsu:2012uh} (green) at Belle, $K\to \pi\slashed{E}$~\cite{Artamonov:2008qb} (blue) at E949, and invisible Higgs decays $h\to \rm inv$ (grey) at the LHC~\cite{Sirunyan:2018koj}. Dashed lines show projections for rare meson decays at Belle II (yellow, green, red) and NA62 (blue), and for $h\to \rm inv$ (light grey) and the Higgs signal strength $\mu$ (dark grey) at the HL-LHC. All results correspond to exclusions at $95\%$ CL. The mass and coupling of dark fermions are set to $m_\chi = 0$ and  $y_\chi = 1$.}
	\label{fig:inv}
\end{figure}

Since the dark particles escape the detector, the final state is the same as in $B\to K\nu\bar\nu$ with neutrinos. Searches for $B\to K \nu\bar\nu$ have been performed by BaBar~\cite{delAmoSanchez:2010bk,Lees:2013kla} and Belle~\cite{Lutz:2013ftz,Grygier:2017tzo} using both hadronic and semi-leptonic $B$ tags. However, the derived bounds on the branching ratio $\mathcal{B}(B\to K \nu\bar\nu)$ rely on the three-body kinematics of the standard-model process, so that we cannot reinterpret them for the two-body decays $B\to KS(\to \chi\bar{\chi})$.~\footnote{A similar observation has been made in Ref.~\cite{MartinCamalich:2020dfe} in search of stable axions in $B\to K\slashed{E}$.} BaBar provides model-independent bounds on the $B\to K\slashed{E}$ distribution in bins of the momentum transfer $q^2 = (p_B-p_K)^2$~\cite{Lees:2013kla}. Since the momentum distribution in $B\to K \chi\bar{\chi}$ peaks sharply around the scalar resonance $q^2 = m_S^2$, we combine the three bins with the largest predicted rates and add uncertainties in quadrature. The resulting bounds on the parameter space $\{m_S,\theta\}$ are shown in Fig.~\ref{fig:inv}, excluding mixing angles larger than $\theta \approx 0.006$ at the 95\% CL.

At Belle II, we expect a higher sensitivity to dark scalars due to the much larger data set. A dedicated simulation of $B\to K \nu\bar\nu$ predicts that with 50/ab of data Belle II can measure the branching ratio with about 10\% precision~\cite{Kou:2018nap}. Since the dependence of this prediction on three-body kinematics has been reduced by choosing different selection variables, we can use it to estimate the reach for dark scalars. In Fig.~\ref{fig:inv} we show that Belle II can probe scalars with mixing angles down to $\theta \approx 10^{-3}$ and masses $m_S > m_B - m_K$ beyond the resonance region. To optimize the sensitivity for light resonances in $B\to K\slashed{E}$ and distinguish them from  $B\to K\nu\bar\nu$ background, we suggest to perform a dedicated search for two-body contributions of $B\to KS(\to \chi\bar\chi)$ in $B\to K\slashed{E}$ decays.

Light scalars can also be probed in searches for $K\to \pi \nu\bar\nu$ at fixed-target experiments. The currently strongest bound on $\mathcal{B}(K^+\to \pi^+\slashed{E})$ by E949~\cite{Artamonov:2008qb} excludes the blue region in Fig.~\ref{fig:inv} at 95\% CL. NA62 has the potential to improve the sensitivity by more than a factor of two~\cite{CortinaGil:2018fkc,Lurkin:2019brq}.

In addition to $B\to K\chi\bar\chi$ and $K\to \pi\chi\bar\chi$, dark sectors can induce fully invisible meson decays~\cite{Badin:2010uh,Kamenik:2011vy}. In our model the branching ratio is  ($q=s,d$)
\begin{align}\label{eq:binv}
        \mathcal{B}(B_q\to\chi\bar\chi) & = \frac{f_{B_q}^2 m_{B_q}^5}{32\pi\,\Gamma_{B_q}} \frac{\sqrt{2}G_F\left|C_{bq}\right|^2  y_\chi^2 c_\theta^2}{(m_{B_q}^2-m_S^2)^2 + m_S^2\Gamma_S^2} \\\nonumber
        &\phantom{=}\times\frac{(m_b-m_q)^2}{(m_b+m_q)^2}\bigg(1-\frac{4m_\chi^2}{m_{B_q}^2}\bigg)^{\frac{3}{2}}\,,
\end{align}
where $\Gamma_{B_q}$ and $f_{B_q}$ are the total decay rate and the hadronic decay constant of the $B_q$ meson and we have neglected $m_\chi$ in the last relation. Belle~\cite{Hsu:2012uh} and BaBar~\cite{Lees:2012wv} have searched for invisible $B_d$ decays. The currently strongest upper bound $\mathcal{B}(B_d \to \slashed{E}) < 1.4\times 10^{-5}$~\cite{Hsu:2012uh} excludes dark scalars around the $B_d$ meson mass, as shown in Fig.~\ref{fig:inv}. Unlike resonant $B\to K\chi\bar\chi$ decays, the sensitivity of $B_d\to \chi\bar\chi$ depends on the coupling and mass of the dark fermions, see Eq.~\eqref{eq:binv}. Using projections for Belle II~\cite{Kou:2018nap}, we show that searches for $B_d \to \slashed{E}$ and $B_s \to \slashed{E}$ can extend the reach for invisibly decaying scalars by about an order of magnitude.

Independent bounds on dark scalars can finally be derived from Higgs observables at the LHC~\cite{Barger:2007im,Freitas:2015hsa}. Scalar mixing induces invisible Higgs decays $h\to \chi\bar\chi$ with a branching ratio
\begin{align}
\mathcal{B}(h\to \text{inv}) = \frac{s_\theta^2 \Gamma_{\chi\bar{\chi}}^h}{c_\theta^2 \Gamma_{\rm SM}^h + s_\theta^2 \Gamma_{\chi\bar{\chi}}^h}\,,
\end{align}
where $\Gamma_{\rm SM}^h$ is the total Higgs width in the standard model and $\Gamma_{\chi\bar\chi}^h$ is the partial decay width to dark fermions, defined as in Eq.~\eqref{eq:partial} with $m_S\to m_h$. The current upper bound on invisible Higgs decays, $\mathcal{B}(h\to \rm inv) < 0.22$~\cite{Sirunyan:2018koj} excludes mixing angles larger than $\theta \approx 0.015$ at 95\% CL. Projections for the HL-LHC predict an extended reach to $\theta \approx 0.005$.

Scalar mixing also causes a universal reduction of all Higgs couplings to visible particles by $c_\theta$. This suppresses the Higgs signal strength defined by
\begin{align}
\mu = \frac{\sigma_h \times \mathcal{B}(h\to \rm vis)\phantom{_{\rm SM}}}{\sigma_h\times  \mathcal{B}(h\to \rm vis)_{\rm SM}} = c_\theta^2\frac{c_\theta^2\Gamma_{\rm SM}^h}{c_\theta^2\Gamma_{\rm SM}^h + s_\theta^2\Gamma_{\chi\bar \chi}^h}\,,
\end{align}
where $\sigma_h$ is the Higgs production rate and $\mathcal{B}(h\to \rm vis)$ the branching ratio to visible final states. Current global analyses constrain universal modifications of the Higgs couplings, but without allowing for invisible decays. For the HL-LHC, such an analysis has been performed assuming Run-2 systematics~\cite{Cepeda:2019klc}. The expected reach for dark scalars depends on the invisible decay rate $\Gamma_{\chi\bar\chi}^h$. For $y_\chi = 1$ we expect that mixing angles down to $\theta \approx 0.008$ will be probed. The sensitivity is comparable with the current BaBar bounds from $B\to K\slashed{E}$, but less than predicted at Belle II.

\section{Displaced vertex signatures}\label{sec:dv}
\noindent
If invisible decays are kinematically forbidden or absent, dark scalars leave signatures with visible decay products. Due to the flavor-hierarchical couplings, scalar decays to light leptons or mesons are suppressed, while scalar production through the top-quark coupling is sizeable even for small mixing $\theta$. The scalar has a nominal lifetime of roughly $c\tau_S = c/\Gamma_S \approx s_\theta^{-2}\,\rm nm$ and becomes long-lived at detector scales for $\theta \lesssim 10^{-2}$. This leads to signatures with displaced vertices, which are perfect targets for flavor or beam dump experiments.

At $e^+e^-$ colliders, light scalars can be abundantly produced from $B\bar{B}$ pairs at the $\Upsilon(4S)$ resonance with subsequent $B\to K S$ decays. Direct production via $e^ +e^-\to S$ is strongly suppressed by the tiny electron coupling. Alternative searches for radiative Upsilon decays $\Upsilon(n) \to S \gamma$ through the $b$-quark coupling at BaBar exclude strong mixing $\theta \gtrsim 0.1$~\cite{Lees:2011wb,Lees:2012iw,Lees:2012te}.

\begin{figure}[t!]
	\centering
	\includegraphics[width=1.02 \linewidth]{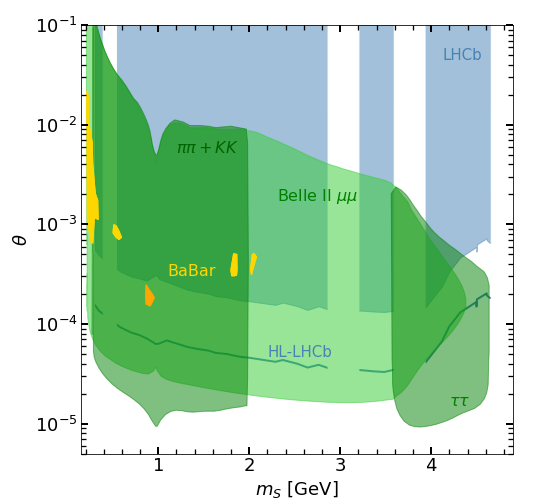}
	\caption{Searches for dark scalars with displaced vertices at flavor experiments. Shown are 95\% CL bounds from $B^+\to K^+S(\to \mu\bar\mu)$ searches at LHCb~\cite{Aaij:2016qsm} (blue) and 90\% CL bounds on $\mathcal{B}(B\to X_s S)\mathcal{B}(S\to f)$ with $f=\mu^+\mu^-$ (yellow) and $\pi^+\pi^-$ (orange) from an inclusive search by BaBar~\cite{Lees:2015rxq}. Regions with 3 or more signal events at Belle II with 50/ab are shown for $B\to KS(\to f)$ with $f=\pi^+\pi^- + K^+K^-,\,\mu^+\mu^-$ and $\tau^+\tau^-$ (green). For comparison, we show projections for $B\to K \mu\bar\mu$ for the high-luminosity phase of LHCb (blue curve).}
	\label{fig:displaced}
\end{figure}

Measurements of $B\to K^{(\ast)}\mu\bar\mu$ decays by BaBar, Belle and LHCb exclude scalar mixing down to $\theta \approx 10^{-3}$~\cite{Schmidt-Hoberg:2013hba}. The event selection is typically restricted to prompt decays. LHCb has performed dedicated searches for displaced muons from long-lived scalars~\cite{Aaij:2015tna,Aaij:2016qsm}. By reinterpreting the search for $B^+\to K^+S(\to\mu\bar\mu)$~\cite{Aaij:2016qsm} we exclude scalar mixing down to $\theta \approx 10^{-4}$, shown in blue in Fig.~\ref{fig:displaced}.
Vetoed regions around the resonances $K_S^0$, $\psi(2S)$ and $\psi(3770)$ are partially excluded by a similar search for $B^0\to K^\ast S(\to\mu\bar\mu)$ decays~\cite{Aaij:2015tna}.

To date, the only search for long-lived scalars at $e^+e^-$ colliders is an inclusive search for displaced vertices of charged leptons, pions or kaons by BaBar~\cite{Lees:2015rxq}. From this analysis BaBar has derived upper bounds on the branching ratio $\mathcal{B}(B\to X_s S)\mathcal{B}(S\to f)$ for different final states $f$. In Fig.~\ref{fig:displaced} we show our reinterpretation of these bounds for $f=\mu^+\mu^-$ (yellow) and $f=\pi^+\pi^-$ (orange). The sensitivity is limited by hadronic backgrounds from $K_S^0$, $\Lambda$, $K^\pm$ and $\pi^\pm$ decays and by the available data set, so that only a few small parameter regions can be excluded.

The fact that BaBar probes very small mixing without optimizing their analysis for dark scalars suggests that Belle II can reach a better sensitivity with a dedicated search. We suggest to search for displaced vertices from exclusive $B\to K S(\to f)$ decays at Belle II, where $K$ stands for either $K^0$, $K^+$, or $K^\ast$ excitations. Promising final states are $f = \mu^+\mu^-$ and $\pi^+\pi^-$, $K^+ K^-$ for scalar masses $m_S \lesssim 2\,\text{GeV}$, as well as $\tau^-\tau^+$, $D^+ D^-$ or $4\pi$ for heavier scalars.

Let us first focus on displaced muon pairs, which probe a large range of scalar masses $2m_\mu < m_S < m_B - m_K$. The signal is defined by a displaced muon vertex and a kaon, which together reconstruct the $B$ momentum. The reconstruction of the full final state strongly suppresses background from displaced hadron decays, making this search largely background-free.

The number of observable displaced muon pairs at Belle II strongly depends on the detector geometry and on the boost of the dark scalar in $B\to K S$ decays. For a good reconstruction efficiency of a displaced vertex, the scalar should decay at a radius larger than vertex resolution $\rho_{\rm min}=500\,\mu \rm m$ and within the Central Drift Chamber (CDC), whose outer radius lies at $\rho_{\rm max} = 113\,\rm cm$ around the beam pipe. The CDC covers the angular range between $\vartheta_{\min} = 17^\circ$ and $\vartheta_{\max} = 150^\circ$~\cite{Kou:2018nap}. The expected number of muon pairs produced within this detector region is
\begin{align}
N_{\mu\bar\mu} &= N_{B\bar{B}} \times 1.93\,\mathcal{B}(B \to K S) \mathcal{B}(S \to \mu\bar\mu)\\\nonumber
& \quad\times \frac{1}{2}\int d\vartheta_0\, \frac{\sin\vartheta_0}{d_S}\int dr\,e^{- \tfrac{r}{d_S}},
\end{align}
where $r$ is the radial distance from the $e^+e^-$ interaction point. The angle $\vartheta_0$ describes the scalar's polar momentum direction in the rest frame of the $B$ meson. Whether or not the scalar decays within the CDC crucially depends on its decay length in the lab frame, $d_S$. Since the distribution of the scalar in $B\to KS$ is spherically symmetric, both $\vartheta_0$ and $d_S$ can be expressed in terms of $\vartheta$ and the boost of the $B$ meson in the lab frame. In our calculations we use the average $B$ boost, $\langle\gamma_B\rangle \approx 1.04$. The expected number of $B\bar B$ pairs in $50/\text{ab}$ of data is $N_{B\bar{B}} = 5\times 10^{10}$, where $B$ denotes the sum of produced $B^+$ and $B^0$ mesons. The factor of $1.93$ takes account of the different lifetimes of $B^+$ and $B^0$. We sum over all relevant $B\to K S$ decay channels with neutral and charged kaons, including $K$, $K^\ast_{700}$, $K^\ast_{892}$, $K^\ast_{1270}$, $K^\ast_{1400}$, $K^\ast_{1410}$, $K^\ast_{1430}$, and $K^\ast_{1680}$. For the pseudo-scalar kaons $K$ we use $B\to K$ hadronic form factors with input from lattice calculations~\cite{Bailey:2015dka}, for the vector resonance $K^\ast_{892}$ we use input from light-cone sum rules~\cite{Gubernari:2018wyi}, and for all other resonances we adopt the approach of Ref.~\cite{Boiarska:2019jym}. In App.~\ref{sec:app2} we give more details about the impact of different kaon final states on the sensitivity to long-lived scalars.

In Fig.~\ref{fig:displaced} we show the parameter region with at least $N_{\mu\bar\mu} = 3$ displaced muon pairs from $B\to K S(\to \mu\bar\mu)$ decays at Belle II (the light green area). The displayed region corresponds to a rejection of the background-only hypothesis at $95\%\ \text{CL}$, assuming Poisson statistics, a reconstruction efficiency $\epsilon$ of $100\%$ and no observed background events.~\footnote{A reduced efficiency would decrease the significance in $\theta$ by roughly $\sqrt{\epsilon}$.} The sensitivity extends between $10^{-2} > \theta > 10^{-5}$, where the upper bound on the mixing is determined by how many scalars decay at $\rho > \rho_{\rm min}$. The lower bound is mostly determined by the low event rate and for light scalars also by how many decays occur at $\rho < \rho_{\rm max}$. The sensitivity to small mixing angles is higher for heavier scalars, which have a shorter decay length and are more likely to decay within the CDC.

The presence of background affects our results only mildly. For instance, by assuming three or less background events the sensitivity is reduced by at most a factor of two in $\theta$. Systematic uncertainties on the signal and/or background could affect the sensitivity further, but can only be determined in a dedicated experimental analysis.

For $\theta < 10^{-5}$ the scalars decay mostly outside the CDC. Their decay products could still be detected by the electromagnetic calorimeter around the CDC, which however is not optimized for vertex reconstruction. Extending the search to the calorimeter can lead to a slightly enhanced sensitivity for $m_S \lesssim 1\,\text{GeV}$.

Searches for other decay channels can further enhance the sensitivity to scalars with small mixing. In Fig.~\ref{fig:displaced} we show the 3-event regions for the final states $\pi^+\pi^- + K^+K^-$ and $\tau^+\tau^-$, obtained as for the muon channel. Pions increase the reach at low masses, especially around $m_S \approx 1\,\text{GeV}$. Here the decay width of the scalar is dominated by the resonance $f_0(980)$, yielding larger event rates in the outer region of the CDC. Adding kaon final states enhances the sensitivity to small $\theta$ by about a factor of three for $m_S \gtrsim 1\,\text{GeV}$. Scalars in the range $2\,\text{GeV} \lesssim m_S < 2 m_\tau$ decay mostly into final states with multiple hadrons. Since calculations of the partial decay widths are very challenging and subject to large uncertainties~\cite{Winkler:2018qyg}, we do not show predictions for hadronic channels in this region. Above the production threshold $m_S > 2m_\tau$, scalar decays to tau lepton pairs dominate over muons due to the flavor-hierarchical coupling. Searches for $\tau^+\tau^-$ extend the sensitivity to heavy scalars significantly. We expect a similar reach for $D^+D^-$ pairs, keeping in mind that for $m_S \gtrsim 2m_\tau$ the sensitivity to $\tau^+\tau^-$ and $D^+ D^-$ final states is subject to hadronic uncertainties due to the presence of charmonium resonances~\cite{Winkler:2018qyg}. If invisible decays are open, the lifetime of the scalar is reduced and the branching ratio to visible final states decreases. The impact on displaced signatures is mild, unless $S\to \chi\bar{\chi}$ decays dominates the scalar width.

Compared with BaBar and Belle, the high search potential at Belle II is mostly due to the larger event rates. The total expected luminosity at Belle II is 30 times larger, which enhances the sensitivity in $\theta$ by about a factor of $\sqrt{30}$.

Compared with searches for $B\to K S(\to \mu\bar{\mu})$ at LHCb, Belle II can probe dark scalars with smaller mixing and larger lifetimes. The main reason is the large boost of $B$ mesons produced at LHCb, so that only scalars with a nominal decay length $c\tau_S \lesssim 30\,\rm cm$ decay inside the vertex detector. With $300/\rm fb$ of data expected during the high-luminosity phase of the LHC, LHCb can extend its reach in $\theta$ by about a factor of four. To obtain the projection shown in Fig.~\ref{fig:displaced}, we have rescaled the number of produced $B$ mesons with the higher luminosity and assumed a background-dominated analysis for $c\tau_S<10\,\text{ps}$ and a background-free search in the region with displaced decays $c\tau_S>10\,\text{ps}$, following Ref.~\cite{Bondarenko:2019vrb}. At Belle II the $B$ mesons are slow and scalars decay within the CDC for $c\tau_S \lesssim 10\,\rm m$. 

\begin{figure}[t!]
 	\centering
 	\includegraphics[width=1.02 \linewidth]{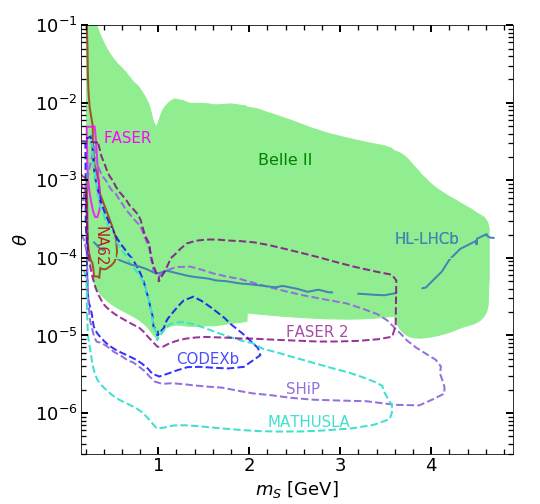}
 	\caption{Long-lived dark scalars at Belle II and LHCb versus long-baseline experiments. Shown is the sensitivity to a displaced vertex signal for the combined channels $\mu^+\mu^-$, $\pi^+\pi^-$, $K^+K^-$ and $\tau^+\tau^-$ at Belle II (green), as well as sensitivity projections for running or funded experiments (plain curves) and for proposed future experiments (dashed curves).}
 	\label{fig:future}
 \end{figure}

Searches for long-lived scalars in rare $B$ decays at Belle II are complementary to searches for similar processes at fixed-target experiments. In Fig.~\ref{fig:future} we show the predicted sensitivity of future searches for displaced muons at NA62 in beam-dump mode~\cite{Winkler:2018qyg,Lanfranchi:2017wzl}, FASER~\cite{Ariga:2018uku} and FASER~2~\cite{Feng:2017vli,Ariga:2018uku}. Remarkably, Belle II has a similar sensitivity to scalar mixing with a much shorter baseline. Other dedicated proposed experiments for long-lived particles like SHiP~\cite{Alekhin:2015byh}, CODEX-b~\cite{Gligorov:2017nwh} or MATHUS\-LA~\cite{Evans:2017lvd} could extend the reach to scalars with even smaller mixing.

\section{Conclusions}\label{sec:conclusions}
\noindent
We have proposed a new search for dark scalars in displaced $B\to KS(\to f)$ decays at Belle II. We predict a high sensitivity to scalars in the GeV range with mixing angles in the range of $10^{-2} > \theta > 10^{-5}$, which is mostly due to the large data set and the low $B$ meson boost at Belle II. The results can be interpreted for a large class of models with a light scalar singlet, relying only on the scalar's lifetime and mixing with the Higgs field. In particular, Belle II can search for thermal relics where the scalar acts as a mediator to a heavier dark matter candidate~\cite{Krnjaic:2015mbs,Bondarenko:2019vrb}. Complementary searches for invisibly decaying scalars in $B\to K\slashed{E}$ and $B_q\to \slashed{E}$ probe mixing angles of $\theta \approx 10^{-3}$, comparable to invisible Higgs decays at the LHC. We look forward to Belle II's first results on light dark scalars.

\section{Acknowledgments}
\noindent
We thank Florian Bernlochner, Torben Ferber, Christopher Smith, Phillip Urquijo, Mike Williams and Robert Ziegler  for helpful discussions and Martin Winkler for providing us with numerical results for the hadronic scalar decay rates. We are grateful to the authors of Ref.~\cite{Kachanovich:2020yhi} for pointing out a mistake in our calculation of the Belle II detector geometry, which we have corrected in this revised version. AF and RS acknowledge support of the DFG (German Research Foundation) through the research training group \emph{Particle physics beyond the Standard
Model} (GRK 1940). The research of SW is supported by the Carl Zeiss foundation through an endowed junior professorship (\emph{Junior-Stiftungsprofessur}) and by the DFG under grant no.  396021762--TRR 257.  


\appendix
\section{Scalar Higgs portal}\label{sec:app}
\noindent In our scenario, the scalar singlet $\phi$ mixes with the neutral component, $h_0$, of the SM Higgs doublet $H = (h_+,(v+h_0+i\eta_0)/\sqrt{2})^\top$ through electroweak symmetry breaking. The corresponding mixing angle $\theta$ is defined by
\begin{align}\label{eq:mixing}
	\sin^2\theta & =\frac{1}{2}\bigg(1+\frac{m_\phi^2-m_{h_0}^2}{\Delta m^2}\bigg)\,,\\\nonumber
   (\Delta m^2)^2 & = 4v^2\lambda_3^2+\big(m_\phi^2-m_{h_0}^2\big)^2\,. 	
\end{align}
The masses of the physical states $S$ and $h$ are 
\begin{align}\label{eq:mass-es}
    m_S^2 & = \tfrac{1}{2}\left(m_\phi^2-m_{h_0}^2 - \Delta m^2\right)\,,\\\nonumber
	m_h^2 & = \tfrac{1}{2}\left(m_\phi^2-m_{h_0}^2 + \Delta m^2\right) \approx (125\,\text{GeV})^2\,.
\end{align}
\section{Combining different kaon channels}\label{sec:app2}
\begin{figure}[h!]
 	\centering
 	\includegraphics[width=1.02 \linewidth]{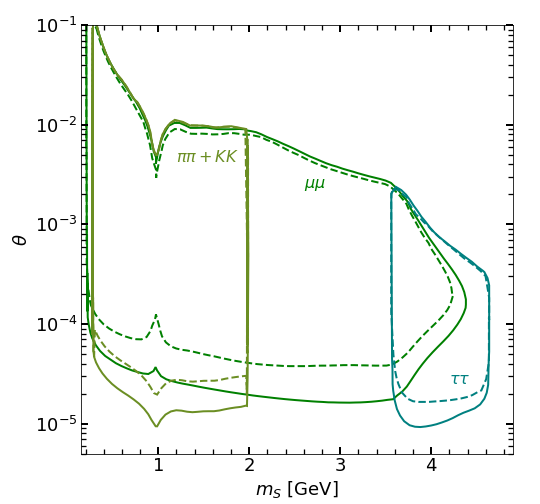}
 	\caption{Sensitivity to $B\to KS$ decays with different kaons in the final state. Plain curves: combination of 8 kaon resonances. Dashed curves: pseudo-scalar kaons $K^{\pm}$ only.}
 	\label{fig:kaons}
 \end{figure}

\noindent The sensitivity to long-lived scalars is enhanced by combining searches for $B\to KS$ decays with different kaons in the final state. To quantify this effect, we compare two different scenarios in Fig.~\ref{fig:kaons}: the sum of 8 kaon resonances $K$, $K^\ast_{700}$, $K^\ast_{892}$, $K^\ast_{1270}$, $K^\ast_{1400}$, $K^\ast_{1410}$, $K^\ast_{1430}$, and $K^\ast_{1680}$, including both charged and neutral kaons (plain curves), and $K^{\pm}$ only (dashed curves). By adding higher kaon resonances the sensitivity to small mixing angles is enhanced by a factor of two or more in most of the parameter region.

\bibliography{dark-sectors}

\end{document}